\documentclass[11pt]{article}
\usepackage{amsmath}
\usepackage{graphicx,psfrag,epsf}
\usepackage{enumerate}
\usepackage{natbib}
\usepackage{url}
\usepackage{caption,subcaption}
\usepackage{xfrac}
\usepackage{multirow}
\usepackage{amssymb}
\usepackage{mathtools}

\usepackage[affil-it]{authblk}
\usepackage[margin=1in]{geometry}

\usepackage[parfill]{parskip}
\usepackage{indentfirst}
\usepackage[defaultlines=2,all]{nowidow}

\begin{document}

\title{Computation of Two and Three Dimensional Confidence Regions with the Likelihood Ratio}
\author{Adam Jaeger \thanks{Statistical and Mathematical Sciences Institute, Research Triangle Park NC, \texttt{ajaeger@samsi.info}} 
\thanks{Department of Statistical Sciences, Duke University, Durham NC, \texttt{apjaeger@stat.duke.edu}}}
\date{}

\maketitle

\begin{abstract}
The asymptotic results pertaining to the distribution of the log likelihood ratio allow for the creation of a confidence region, which is a general extension of the confidence interval.  Two and three dimensional regions can be displayed visually in order to describe the plausible region of the parameters of interest simultaneously.  While most advanced statistical textbooks on inference discuss these asymptotic confidence regions, there is no exploration of how to numerically compute these regions for graphical purposes.  This article demonstrates the application of a simple trigonometric identity to compute two and three dimensional confidence regions; we transform the Cartesian coordinates to create what we call the radial profile log likelihood.  The method is applicable to any distribution with a defined likelihood function, so it is not limited to specific data distributions or model paradigms. We describe the method along with the algorithm, follow with an example of our method and end with an examination of computation time. 
\end{abstract}
\section{Introduction}
A confidence region is a high dimensional generalization of a confidence interval; it describes the $100(1-\alpha)$\% confidence area of a multi-dimensional parameter.  Unlike family wise error corrections of simultaneous confidence intervals, confidence regions account for the probabilistic relationship between the variables, resulting in a more precise description of the confidence bounds for the parameters simultaneously. 

The limiting distribution of the log likelihood ratio statistic, which can be used to create asymptotic confidence regions, is discussed in textbooks commonly used in a mathematical statistics class \citep[such as][]{casellaberger2002,wms2008}. One method to graphically display a desired region is to compute the log likelihood ratio over a grid of parameter values, and then estimate the bound from these values. This approach is sufficient for most cases, but there are two major drawbacks.  First, if the data are highly variable the number of values to compute becomes very large, having a significant impact on computation time. Second, if the range of parameter values is not properly selected the confidence region may not be captured.

We demonstrate a method for computing two and three dimensional confidence regions that is applicable to any case where the likelihood function can be expressed. Our method uses a basic trigonometric identity allowing us to greatly simplify the computation of the confidence bound, and eliminates the issue of specifying a range of parameter values.

\section{Confidence Regions Utilizing Likelihood Ratio Test}

We start with a brief summary of likelihood functions and the likelihood ratio.  For the remainder we will denote random variables with upper case letters. Sample values will be denoted with lower case letters.

Let $X_i$ be a random variable from some distribution $f_X$ with $p\times 1$ parameter vector $\theta$.  Given a simple random sample $x_1,\ldots,x_n$  the likelihood function for $\theta$ is 
\begin{align*}
\mathcal{L}(\theta) =  \mathcal{L}(\theta;x)  =   \prod_{i=1}^n f_X(x_i;\theta).
\end{align*}
The value $\hat{\theta} = \hat{\theta}(x_1,\ldots,x_n)$  that maximizes $\mathcal{L}(\theta)$ is called the maximum likelihood estimator (MLE) of the true parameter vector $\theta_0$; specifically for a sample $x_1, \ldots, x_n$ the maximum likelihood estimator is
\begin{align*}
\hat{\theta} = \sup_\theta \prod_{i=1}^n f_X(x_i;\theta).
\end{align*}
It should be noted that MLEs do not always exist and may not be unique. We will assume for the remainder that unique MLEs exists, and as a consequence
\begin{align*}
\mathcal{L}(\hat{\theta}) > \mathcal{L}(\theta) \quad \mbox{ for all } \theta \neq \hat{\theta}.
\end{align*}

In most cases it is computationally advantageous to work with the the natural log of the likelihood function, 
\begin{align*}
	\ell(\theta) = \ell(\theta;x)   =   \log\mathcal{L}(\theta) = \sum_{i=1}^n \log f_X(x_i;\theta)
\end{align*}
which is referred to as the log likelihood.  The log likelihood ratio 
\begin{align}
T(\theta) = -2 \log\left( \frac{\mathcal{L}({\theta})}{\mathcal{L}(\hat{\theta})}\right) = -2[\ell({\theta}) - \ell(\hat{\theta}) ] \label{eqn:llT}
\end{align}
is approximately $\chi^2(p)$ when $n$ is large \citep{wilks1938}. This gives rise to the likelihood ratio test, which we can use to compute the 
$100(1 - \alpha)$\% confidence bounds. 

In many cases we have what are commonly referred to as nuisance parameters.  These are parameters that must be estimated in order to compute the likelihood function but are not of interest for the analysis.  Let  $\theta$ denote the $p \times 1$ subset of parameters we are interested in, and $\nu$ denote the $q \times 1$ set of nuisance parameters. The profile log likelihood function is
\begin{align*}
	\hat{\ell}(\theta) = \sup_\nu \ell(\theta,\nu) = \ell(\theta,\hat{\nu}(\theta)),
\end{align*}
where $\hat{\nu}(\theta)$ is the value that maximizes the log likelihood function given the value of $\theta$ \citep[][pages 61-62]{pawitan2013}. Since $\hat{\nu}(\theta)$ adds additional computation a common approach is to replace $\hat{\nu}(\theta)$ with $\hat{\nu}(\hat{\theta}) = \hat{\nu}$, which is simply the MLE of the nuisance parameters.  This substitution is theoretically justified when the maximum likelihood estimator is consistent \citep[][pages 90-92]{barndorffnielsencox1994}.  We now have the profile log likelihood 
\begin{align*}
	\hat{\ell}(\theta) = \ell(\theta,\hat{\nu}),
\end{align*}
and the profile log likelihood ratio 
\begin{align}
T(\theta) = -2[\hat{\ell}(\theta)  - \hat{\ell}(\hat{\theta})] \label{eqn:pllT}
\end{align}
is approximately $\chi^2(p)$ when $n$ is large. For the remainder of the article we will work with profile log likelihoods, but when there are no nuisance parameters Equation \ref{eqn:llT} replaces Equation \ref{eqn:pllT}.

We now give a formal definition for the asymptotic $p$-dimensional confidence bound. The boundary of a $100(1-\alpha)$\% confidence region is the set
\begin{align*}
B_\theta = \left\{\forall \theta \in \mathbb{R}^p \left| T(\theta) = \chi^2_{(1 - \alpha)}(p)  \right. \right\}
\end{align*}
where $\chi^2_{(1 - \alpha)}(p)$ is the $(1 - \alpha)$ quantile of a chi squared random variable with $p$ degrees of freedom.

\section{The Radial Profile Log Likelihood Ratio}
\label{sec:alg}

\subsection{Two Parameter Confidence Region}
\begin{figure}[ht]
\begin{center}
\begin{tabular}{c}
	\includegraphics[scale=.4]{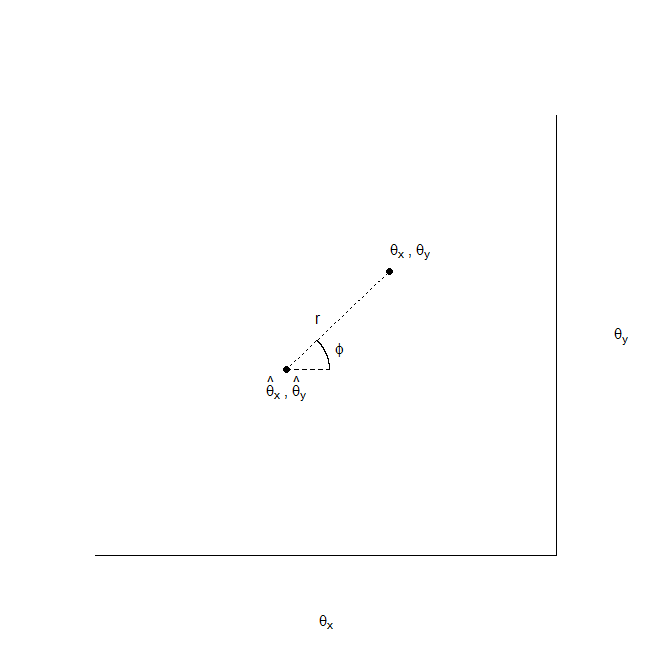}
		\end{tabular}\caption{Visual example of defining a pair $(\theta_x,\theta_y)$ in terms of $(\phi,r)$.} \label{fig:f3}
\end{center}
\end{figure}

The radial profile log likelihood ratio is based on recognizing that any pair of Cartesian coordinates $(x,y)$ on $\mathbb{R}^2$  can be expressed in terms of an angle $\phi$ and a distance $r$ from some origin point. Let the two parameters of interest be the scalars $\theta_x$ and $\theta_y$, with the subscripts denoting the axis the parameter is displayed.  Let $\phi \in [0,2\pi)$ denote the angular coordinate  and  $r \in [0,\infty)$ denote the radial coordinate. Setting the MLEs $\hat{\theta}_x$ and $\hat{\theta}_y$ as the origin, for a fixed $\phi$ and $r$ there exists a $(\theta_x,\theta_y)$ pair
\begin{align}
\begin{split}
\theta_x &= \hat{\theta}_x + r\cos(\phi), \mbox{ and}  \label{eqn:conv}\\
\theta_y &= \hat{\theta}_y + r\sin(\phi).
\end{split}
\end{align} 
Figure \ref{fig:f3} shows a visual representation of Equation \ref{eqn:conv}.

\clearpage

The profile log likelihood for a fixed $\phi$ is
\begin{align*}
\hat{\ell}_\phi(r) &= \hat{\ell}(\hat{\theta}_x + r\cos(\phi),\hat{\theta}_y + r\sin(\phi)),
\end{align*}
which leads to the radial profile log likelihood ratio
\begin{align}
T_\phi(r) = -2(\hat{\ell}_\phi(r) - \hat{\ell}_\phi(0)), \label{eqn:polarT}
\end{align}
and the distance from the MLEs to the boundary of a two parameter $100(1-\alpha)$\% confidence region for a given $\phi$ (which is the radial coordinate) is
\begin{align}
\min_r \left\{r\in\mathbb{R}^1  \left| T_\phi(r) = \chi^2_{1-\alpha}(2)    \right.  \right\}. \label{eqn:cibnd}
\end{align}
There may be cases where there is no solution to Equation \ref{eqn:cibnd}, indicating that the boundary edge extends beyond the parameter space. In these instances $r$ is the maximum value such that the solutions to Equation \ref{eqn:conv} are on the parameter space.

A set of points defining the edge of the confidence region is computed by choosing a set of $\phi$, then for each $\phi$ find the distance $r$ from the MLEs to the boundary edge using Equation \ref{eqn:cibnd}, which is then back transformed to the original Cartesian coordinates using Equation \ref{eqn:conv}.

The most obvious advantage to this method is at each step the points defining the boundary edge are found, so there are not any `wasted' computations.  Furthermore the user does not have to specify any lower or upper bounds for the parameters of interest since each step finds the edge points; instead of having to define a grid of values we simply need to choose how many points we want to use to create the confidence bound. Determination of $r$ for each $\phi$ can be accomplished using a numerical single root solution or a constrained optimization function.

\subsection{Three Parameter Confidence Region}

Extension of the radial profile log likelihood to a three parameter region is done using the spherical coordinate conversion. Define the azimuth $\phi \in [0,2\pi)$, the inclination $\tau \in [0,\pi]$ and radial coordinate $r \in [0,\infty)$. Using the MLEs as the origin the spherical coordinates are 
\begin{align*}
\theta_x &= \hat{\theta}_x + r\cos(\phi)\sin(\tau),\\
\theta_y &= \hat{\theta}_y + r\sin(\phi)\sin(\tau),\\
\theta_z &= \hat{\theta}_z + r\cos(\tau),
\end{align*}
the resultant radial profile log likelihood is
\begin{align*}
\hat{\ell}_{\phi,\tau}(r) &= \hat{\ell}(\hat{\theta}_x + r\cos(\phi)\sin(\tau),\hat{\theta}_y + r\sin(\phi)\sin(\tau), \hat{\theta}_z + r\cos(\tau)),
\end{align*}
the radial profile log likelihood ratio is
\begin{align*}
T_{\phi,\tau}(r) = -2(\hat{\ell}_{\phi,\tau}(r) - \hat{\ell}_{\phi,\tau}(0)),
\end{align*}
and finally the distance to the boundary is 
\begin{align*}
\min_r \left\{r\in\mathbb{R}^1  \left| T_{\phi,\tau}(r) = \chi^2_{1-\alpha}(3)    \right.  \right\}. 
\end{align*}
The steps to compute the region would be the same approach shown using polar coordinates except each $r$ is based on a pair $(\phi,\tau)$.  Like the polar coordinate conversion this also guarantees that all computations define a boundary edge of the region.

\subsection{Example} \label{sec:ex}
We demonstrate how to compute a 90\% confidence region for bivariate normal data using the radial profile log likelihood ratio.  Let $x_1,\ldots,x_n$ be a sample from a bivariate normal distribution. If we are interested in inference on the mean vector $\mu = (\mu_x,\mu_y)^T$ the profile log likelihood ratio is
\begin{align}
T(\mu_x,\mu_y) &= 
	 \frac{n}{1 - \hat{\rho}^2 } \left[ \left( \frac{\hat{\mu}_x - \mu_x}{\hat{\sigma}_x}\right)^2 + \left( \frac{\hat{\mu}_y - \mu_y}{\hat{\sigma}_y}\right)^2 - 2\hat{\rho} \left(\frac{\hat{\mu}_x - \mu_x}{\hat{\sigma}_x} \right)  \left(\frac{\hat{\mu}_y - \mu_y}{\hat{\sigma}_y} \right) \right]\label{eqn:llnorm}
\end{align}
where $\hat{\mu}_x$, $\hat{\mu}_y$, $\hat{\sigma}^2_x$, $\hat{\sigma}^2_y$, and $\hat{\rho}$  are the MLEs.  We then rewrite the log likelihood ratio statistic in terms of $r$ and $\phi$ as shown in  Equation \ref{eqn:polarT}. With some simple algebra we have the radial profile log likelihood ratio is
\begin{align*}
T_\phi(r) &=   \frac{n}{1 - \hat{\rho}^2 } \left[ \left(\frac{r\cos(\phi)}{\hat{\sigma}_x}\right)^2 + \left(\frac{r\sin(\phi)}{\hat{\sigma}_y}\right)^2 - 2\hat{\rho} \left(\frac{r\cos(\phi)}{\hat{\sigma}_x} \right)  \left(\frac{r\sin(\phi)}{\hat{\sigma}_y} \right)                  \right],
\end{align*}
and the distance from center to the edge of the confidence region is
\begin{align*}
\min_r \left\{r\in\mathbb{R}^1  \left| T_\phi(r) = 4.6055    \right.  \right\}. 
\end{align*}
One quick note is since this is normally distributed data we could replace $\chi^2_{1-\alpha}(2)$ with $\frac{2(n-1)}{n-2}F_{1-\alpha}(2,n-2)$ \citep[][page 155]{hardlesimar2007}, but we use the chi squared distribution in keeping with the asymptotic result shown in \citet{wilks1938}.

The algorithm is as follows: let $\phi_1,\ldots,\phi_N$ be a set of angles defined on $[0,2\pi)$. For $j=1,\ldots,N$ 
\begin{enumerate}
	\item Determine $r_j$ such that $T_{\phi_j}(r_j) = 4.6055$
	\item Set $\mu_x^{(j)} =\hat{\mu}_x + r_j\cos(\phi) $	and $\mu_y^{(j)} =\hat{\mu}_y + r_j\sin(\phi)$
\end{enumerate}
which will result in $N$ pairs of $\mu_x$ and $\mu_y$ corresponding to the edge of the confidence region.  

To examine the computational time we generate 10 random values from a bivariate normal with $\mu_x = \mu_y = 0$, $\sigma^2_x = \sigma_y^2 = 10$ and $\rho = 0.5$.  We use the radial profile log likelihood with 180 equally spaced angles in $[0,2\pi)$, and for each $\phi$ use the \texttt{uniroot} function in R to solve $T_\phi(r) = 4.6055$. This is compared to the time necessary to compute 14,641 values of Equation \ref{eqn:llnorm} over an evenly spaced set of parameter values in $[-3,3]\times [-3,3]$.

\begin{figure}[ht]	
	\centering
	\begin{subfigure}[t]{.45\textwidth}
		\centering
		\includegraphics[scale = .35]{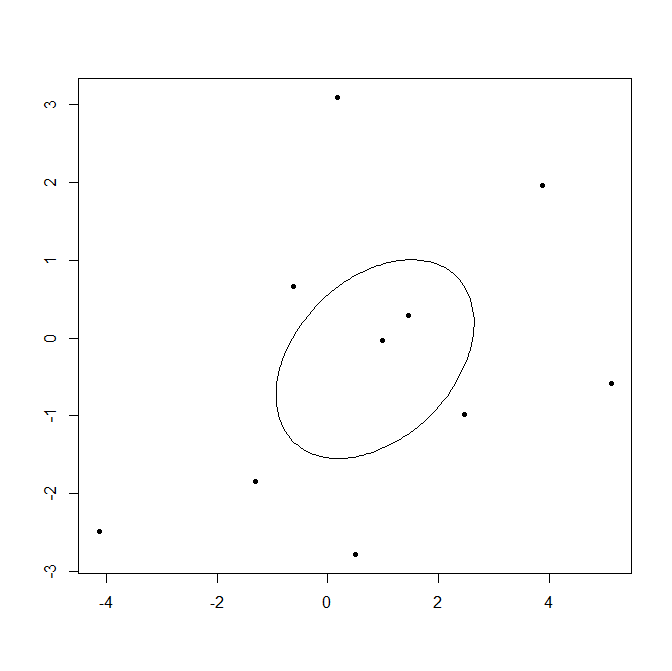}
		\caption{90\% confidence region using radial profile log likelihood ratio. Computation time 0.032 seconds.}\label{fig:f4a}		
	\end{subfigure}
	\quad
	\begin{subfigure}[t]{.45\textwidth}
		\centering
		\includegraphics[scale=.35]{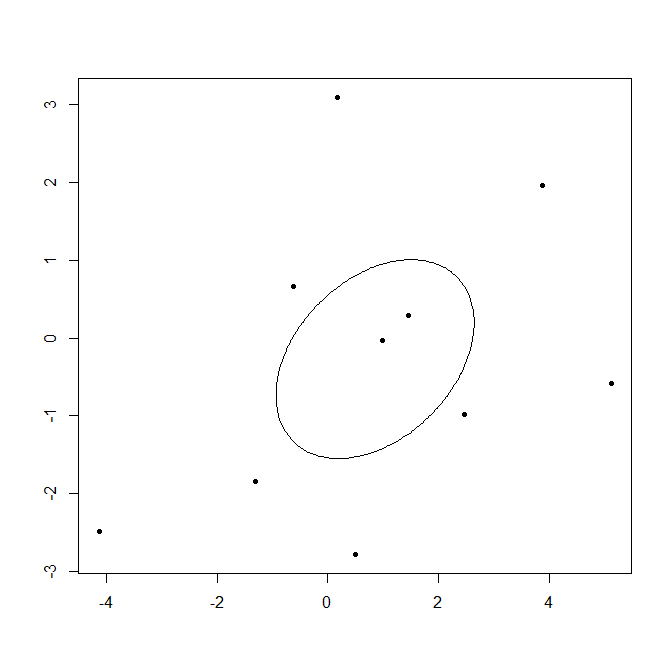}
		\caption{90\% confidence region using grid method. Computation time 0.140 seconds.}\label{fig:f4b}
	\end{subfigure}
	\caption{90\% confidence regions generated from 10 normally distributed observations.}\label{fig:f4}
\end{figure}

Figure \ref{fig:f4} shows that the two methods create graphically indistinguishable regions, but the computational time using the radial profile log likelihood ratio is less.  To demonstrate an even more extreme example we multiply the same data by 10 and compute the 90\% region again, except this time in order to fully capture the domain of the confidence region the parameter values are evenly spaced over $[-30,30]\times [-30,30]$ (total of 1,442,401 computations). The time needed to compute the boundary using the radial profile log likelihood method is 0.031 seconds, while the grid method now requires 142 seconds. Our approach shows a clear computational advantage, being 4,577 times faster. The computation time for the grid method could be reduced by limiting the range of grid values, but that requires a priori knowledge of where the region will fall. Alternatively fewer values over the grid could be used, but this may result in a less precise identification of the region.

\section{Discussion}
The efficiency of the radial profile log likelihood is dependent on the method for determining $r$ along with the choice of how many $\phi$ to examine. From our experience the \texttt{uniroot} function in R results in the fastest computation time, while use of the constrained optimization function \texttt{fmincon} in MatLab requires more time to complete. In regards to the number of angles to use, we have found for graphical purposes 180 is a good lower value.

Although there are cases where using the grid approach would be more efficient than the radial profile log likelihood, the grid method still requires a large number of computations that do not factor into defining the region and may not cover the area defining the boundary of the confidence region.  Given these properties the radial profile log likelihood is especially useful for simulation studies involving confidence regions. Given the computation times seen from the second example, if we had created 100 replicates it would take less than 4 seconds to compute the boundary of the confidence region for all 100 replications. In contrast the grid method would take several hours to accomplish the same task,  and there is still no guarantee that all 100 sets would completely capture the region.

\section*{Acknowledgment}
This material was based upon work partially supported by the National Science Foundation under Grant DMS-1127914 to the Statistical and Applied Mathematical Sciences Institute. Any opinions, findings, and conclusions or recommendations expressed in this material are those of the author(s) and do not necessarily reflect the views of the National Science Foundation. 

\section*{Supplemental Material}
We provide the R code used for both examples from Section \ref{sec:ex}. Additionally we provide R code to run 100 replications of the second example.


\bibliographystyle{apalike}

\bibliography{crbib}

\begin{thebibliography}{}

\bibitem[Barndorff-Nielsen and Cox, 1994]{barndorffnielsencox1994}
Barndorff-Nielsen, O.~E. and Cox, D.~R. (1994).
\newblock {\em Inference and Asymptotics}.
\newblock Chapman \& Hall/CRC, Boca Raton.

\bibitem[Casella and Berger, 2002]{casellaberger2002}
Casella, G. and Berger, R.~L. (2002).
\newblock {\em Statistical Inference}.
\newblock Duxbury Press, Pacific Grove, 2nd edition.

\bibitem[H{\"a}rdle and Simar, 2007]{hardlesimar2007}
H{\"a}rdle, W. and Simar, L. (2007).
\newblock {\em Applied Multivariate Statistical Analysis}.
\newblock Springer-Verlag, Berlin Heidelberg, 2nd edition.

\bibitem[Pawitan, 2013]{pawitan2013}
Pawitan, Y. (2013).
\newblock {\em In All Likelihood : Statistical Modelling and Inference Using
  Likelihood}.
\newblock Clarendon Press, Oxford.

\bibitem[Wackerly et~al., 2008]{wms2008}
Wackerly, D., Mendenhall, W., and Scheaffer, R.~L. (2008).
\newblock {\em Mathematical Statistics with Applications}.
\newblock Duxbury Press, Pacific Grove, 7th edition.

\bibitem[Wilks, 1938]{wilks1938}
Wilks, S.~S. (1938).
\newblock The large-sample distribution of the likelihood ratio for testing
  composite hypotheses.
\newblock {\em The Annals of Mathematical Statistics}, 9(1):60--62.

\end{thebibliography}

\end{document}